\documentclass[conference]{IEEEtran}
\ifCLASSINFOpdf
\else
\fi
\usepackage[cmex10]{amsmath}
\ifCLASSOPTIONcompsoc
  \usepackage[caption=false,font=normalsize,labelfont=sf,textfont=sf]{subfig}
\else
  \usepackage[caption=false,font=footnotesize]{subfig}
\fi
\usepackage{url}

\usepackage{verbatim}
\usepackage{graphicx}
\usepackage{epstopdf}
\usepackage{amsfonts}

\usepackage{enumerate}

\begin{document}
%
\title{Social Interaction in the Flickr Social Network}

\author{\IEEEauthorblockN{Karthik Gopalakrishnan\IEEEauthorrefmark{1}, Arun Pandey\IEEEauthorrefmark{2} and Joydeep Chandra\IEEEauthorrefmark{3}}
\IEEEauthorblockA{Department of Computer Science and Engineering\\
Indian Institute of Technology Patna\\
Patna, India\\
Email: \{\IEEEauthorrefmark{1}karthik.cs11,\IEEEauthorrefmark{2}arun.cs11,\IEEEauthorrefmark{3}joydeep\}@iitp.ac.in}}


%


\maketitle

\begin{abstract}
Online social networking sites such as Facebook, Twitter and Flickr are among the most popular sites on the Web, providing platforms for sharing information and interacting with a large number of people. The different ways for users to interact, such as liking, retweeting and favoriting user-generated content, are among the defining and extremely popular features of these sites. While empirical studies have been done to learn about the network growth processes in these sites, few studies have focused on social interaction behaviour and the effect of social interaction on network growth.

In this paper, we analyze large-scale data collected from the Flickr social network to learn about individual favoriting behaviour and examine the occurrence of link formation after a favorite is created. We do this using a systematic formulation of Flickr as a two-layer temporal multiplex network: the first layer describes the follow relationship between users and the second layer describes the social interaction between users in the form of favorite markings to photos uploaded by them. Our investigation reveals that (a) favoriting is well-described by preferential attachment, (b) over 50\% of favorites are reciprocated within 10 days if at all they are reciprocated, (c) different kinds of favorites differ in how fast they are reciprocated, and (d) after a favorite is created, multiplex triangles are closed by the creation of follow links by the favoriter's followers to the favorite receiver.\\
\end{abstract}


%
\IEEEpeerreviewmaketitle

\section{Introduction}
Online social networking sites such as Facebook, Twitter and Flickr are among the most popular sites on the Web today. Users of these sites connect with each other by becoming friends, followers and so on. They also interact with each other by various means, such as liking, retweeting and favoriting user-generated content. At a network level, the amount of such interaction occurring is enormous: over 1.8 million `likes' were done on Facebook every minute \cite{qmee} in 2013 and the number is likely much higher now. The availability of such data provides an excellent opportunity to analyze interaction dynamics in large-scale social systems. However, few studies \cite{lee2010faving} have looked into the underlying mechanisms of social interaction in these sites. The growth of social networks has also been the subject of a number of research efforts \cite{kossinets2006empirical, kumar2010structure, liben2007link} but the effect of social interaction on network growth is not well understood.\\

In this paper, we take a step in this direction and analyze large-scale data collected \cite{social-cascade-www09} from the online social networking site Flickr, one of the most popular photo-sharing sites on the Web. For our analyses, we systematically formulate Flickr as a two-layer temporal multiplex network: the first layer describes the follow relationship between users and the second layer describes the social interaction between users in the form of favorite markings to photos uploaded by them. We shall henceforth refer to the first layer as the \emph{follow} layer and the second layer as the \emph{social interaction} layer. We examined the differences between snapshots of the social interaction layer at different points in time to learn about its growth and consequently gain an understanding of actual interaction occurring in Flickr. We also examined the differences between snapshots of both layers at different points in time to observe the local effect of each favorite marking occurring in the social interaction layer on link formation in the follow layer.\\

Our investigation of social interaction in Flickr reveals that users create and receive favorites in future in direct proportion to the current number of favorites created and received respectively. Additionally, we find that users explore and create favorites to new users in direct proportion to the current number of users favorited. Further, we find that users attract favorites from new users in direct proportion to the current number of users who have favorited them. We also find that favorites are reciprocated quickly if at all they are reciprocated and observe a difference in reciprocation times for two different kinds of favorites that we define later in this paper. We finally examine the formation of links in the follow layer after a favorite is done in the social interaction layer in a manner that can be termed as \emph{multiplex triangle closure} to establish the importance of considering social interaction with regard to follow layer link formation.\\

We believe our work is an important step towards understanding interaction behaviour in social networks like Flickr and provides useful empirical knowledge for the development of new multiplex growth models for such social networks. The rest of the paper is organized as follows. We introduce the Flickr data set and present our formulation of Flickr as a two-layer multiplex network in Section II. We present empirical analyses to understand the growth of the social interaction layer in Section III and examine link formation by multiplex triangle closure in Section IV. We summarize related work on network growth and social interaction in Section V and conclude in Section VI.

\newpage

\section{Methodology}
In this section, we introduce the Flickr data set and present our formulation of Flickr as a two-layer multiplex network.\\
\subsection{Flickr}
Flickr is a social networking site which at its bare minimum enables users to make friends, share photos and interact by favoriting and commenting on photos. We refer to all the users who a given user follows as the user's friends. Those who follow a given user are termed the user's followers. Those who favorite a user's photo(s) are termed the user's fans.

The data set used in this paper was generated by crawling Flickr with a time granularity of 1 day for a period of 104 consecutive days, recording 2.5 million users, 33 million follow links and 34 million favorite markings over 11 million photos. We refer the readers to \cite{social-cascade-www09} for a detailed description of the data set, the data collection methodology used and the functionalities offered in Flickr.\\

\subsection{Formulation as a Multiplex Network}

\begin{figure}
\centering
\includegraphics[scale=1, width=0.5\textwidth]{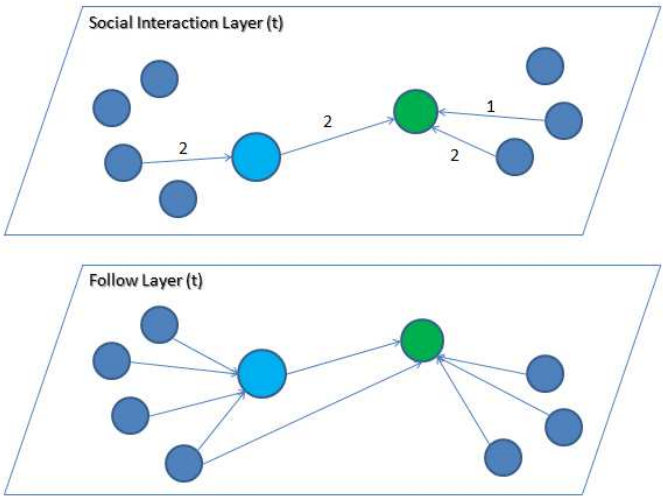}
\caption{A representative multiplex network consisting of the social interaction layer and the follow layer at a given time $t$.}
\label{fig:multiplex}
\end{figure}

Multiplex networks are networks in which a fixed set of nodes are connected by different types of links \cite{boccaletti2014structure}. For our analyses, we view Flickr as a two-layer multiplex network $G^{t} = (G_{f}^{t}, G_{si}^{t})$ at any point in time $t$, where:
\begin{enumerate}[(i)]
\item $G_{f}^{t} = (V, E_{f}^{t})$ is a directed, unweighted graph that represents the follow layer at time $t$.
\item $G_{si}^{t} = (V, E_{si}^{t}, w)$ is a directed, weighted graph that represents the social interaction layer at time $t$.
\item $V$ consists of the entire set of users recorded by the data set.
\item $E_{f}^{t}$ is the edge set of the follow layer as on time $t$, i.e., $e_{ab}^{t} \in E_{f}^{t}$ $\Rightarrow$ as on time $t$, $a$ follows $b$.
\item $E_{si}^{t}$ is the edge set of the social interaction layer as on time $t$, i.e., $e_{ab}^{t} \in E_{si}^{t}$ $\Rightarrow$ as on time $t$, $a$ has favorited $b$'s photos. The number of photos favorited is given by $w(e_{ab}^{t})$, where $w: E_{si}^{t} \to\mathbb N$.\\
\end{enumerate}

Other forms of social interaction such as commenting on photos exist in Flickr. However, they are ignored in our formulation since information such as who commented on a particular photo is not captured by the data set.\\
Note that since the data collection methodology used does not take into account the possible deletion of a favorite marking, $E_{si}^{t_{0}} \subseteq E_{si}^{t_{1}} \subseteq E_{si}^{t_{2}} \subseteq ... \subseteq E_{si}^{t_{n}}$, where $t_{0} \leq t_{1} \leq ... \leq t_{n}$ are different points in time. However, since the methodology does take into account the possible deletion of a follow link, the same cannot be said for the follow edge sets $E_{f}^{t}$ for various points in time.\\

Figure \ref{fig:multiplex} shows a representative multiplex network consisting of the social interaction layer and the follow layer at a given time $t$. We study the growth patterns of the social interaction layer in the next section.\\

\section{Growth of the Social Interaction Layer}
In this section, we empirically examine the growth patterns of the social interaction layer.\\

For our analyses, we first define two kinds of favorites: initiating and continuing. A favorite to a photo uploaded by a given user is termed as an \emph{initiating favorite} if it is the first favorite done by the favoriter to the user. We call it initiating because the favoriter has initiated a directed social interaction relationship with the user through the favorite. A favorite to a photo uploaded by a given user is termed as a \emph{continuing favorite} if it is not the first favorite done by the favoriter to the user. We call it continuing because the favoriter is continuing the directed social interaction relationship previously initiated with the user.\\

\begin{figure*}[!tbh]
\centering \subfloat[$wdeg_{si}^{+}$ (bin) vs. $N_{fav}^{+}$]{\label{fig:pref_fav1}\includegraphics[scale=1, width=0.5\textwidth]{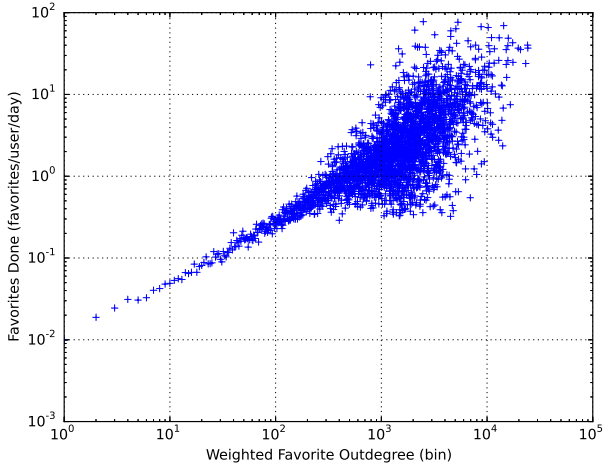}}
\subfloat[$wdeg_{si}^{-}$ (bin) vs. $N_{fav}^{-}$]{\label{fig:pref_rec1}\includegraphics[scale=1, width=0.5\textwidth]{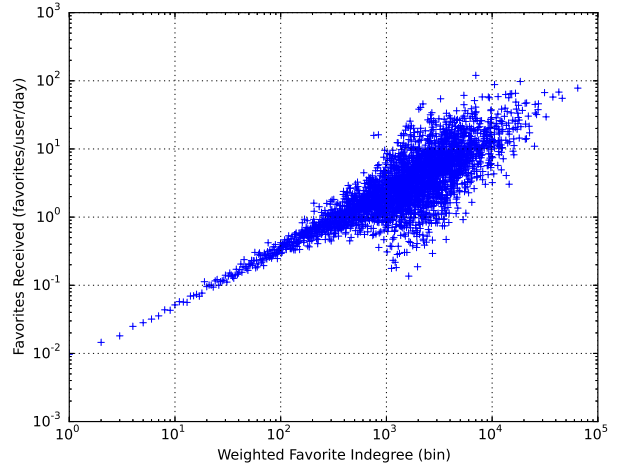}}
\caption{Favoriting of \textit{photos} in Flickr shows evidence of preferential creation [\protect\subref{fig:pref_fav1}] and reception [\protect\subref{fig:pref_rec1}]. $wdeg_{si}^{+}$: weighted favorite outdegree, $N_{fav}^{+}$: number of favorites \textit{created} per day, $wdeg_{si}^{-}$: weighted favorite indegree, $N_{fav}^{-}$: number of favorites \textit{received} per day}
\label{fig:pref1}
\end{figure*}

\begin{figure*}[!tbh]
\centering \subfloat[$deg_{si}^{+}$ (bin) vs. $N_{init}^{+}$]{\label{fig:pref_fav2}\includegraphics[scale=1, width=0.5\textwidth]{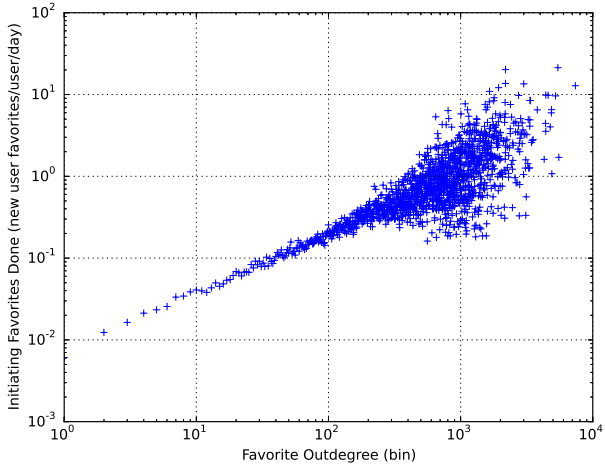}}
\subfloat[$deg_{si}^{-}$ (bin) vs. $N_{init}^{-}$]{\label{fig:pref_rec2}\includegraphics[scale=1, width=0.5\textwidth]{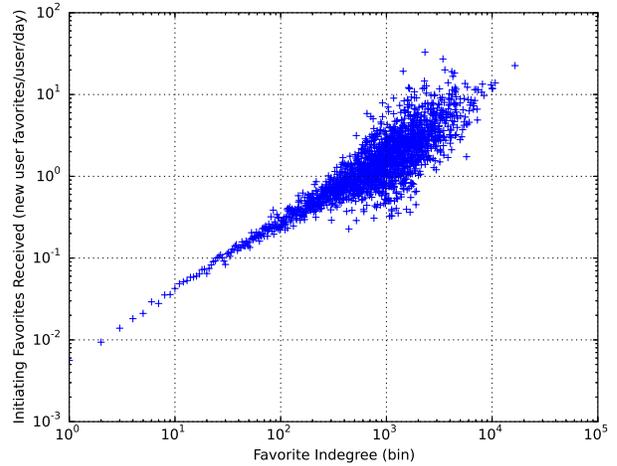}}
\caption{Favoriting of \textit{users} in Flickr shows evidence of preferential creation [\protect\subref{fig:pref_fav2}] and reception [\protect\subref{fig:pref_rec2}]. $deg_{si}^{+}$: favorite outdegree, $N_{init}^{+}$: number of initiating favorites \textit{created} per day, $deg_{si}^{-}$: favorite indegree, $N_{init}^{-}$: number of initiating favorites \textit{received} per day.}
\label{fig:pref2}
\end{figure*}

\subsection{Preferential Attachment}
In this section, we examine preferential attachment in the social interaction layer. Preferential attachment is a network growth phenomenon in which nodes create links preferentially to nodes that already have many links, leading to it also being termed as the `rich get richer' phenomenon. One growth mechanism that follows preferential attachment is the Barabasi-Albert (BA) model \cite{barabasi1999emergence}. Under this model, nodes are selected for new links in linear proportion to their degree.\\

In the case of directed graphs, there can be two kinds of preferential attachment: preferential creation and preferential reception \cite{mislove-2008-flickr}. Nodes create new links in proportion to their outdegree under preferential creation and receive new links in proportion to their indegree under preferential reception. The reasoning given for this split is that users control who they link to but do not have control on who links to them, which is true for the action of favoriting as well. Clearly, both preferential creation and preferential reception can co-occur in a directed graph.\\

From the data set, we constructed $G_{si}^{1}$, the social interaction layer as on the first day of the crawl period and $G_{si}^{104}$, the social interaction layer as on the last day of the crawl period. We first analyzed how the weighted favorite outdegree (the number of favorites created by a node) and weighted favorite indegree (the number of favorites received by a node) in the social interaction layer on the first day correlates with the number of favorites created and received per day until the last day respectively. Figure \ref{fig:pref1} contains plots depicting this analysis: we see that the weighted favorite outdegree positively correlates with the number of favorites created per day (\ref{fig:pref_fav1}), and the weighted favorite indegree positively correlates with the number of favorites received per day (\ref{fig:pref_rec1}). This means that users who favorite a lot of photos continue to favorite a lot of photos, and users whose photos receive a lot of favorites continue to receive a lot of favorites to their photos.

To learn about the tendency of Flickr users to explore the profiles of other Flickr users and favorite their photos (thus increasing the number of users they are fans of), we then analyzed how the actual favorite outdegree (the number of nodes favorited by a node) and favorite indegree (the number of nodes who have favorited a node) in the social interaction layer on the first day correlates with the number of initiating favorites created and received per day until the last day of the crawl period respectively. Figure \ref{fig:pref2} contains plots depicting this analysis: we see that the favorite outdegree positively correlates with the number of initiating favorites created per day (\ref{fig:pref_fav2}), and the favorite indegree positively correlates with the number of initiating favorites received per day (\ref{fig:pref_rec2}). This means that users who explore the profiles and favorite the photos of many users continue to be exploratory and favorite the photos of many more users. Likewise, users whose photos receive favorites from many users continue to receive attention in the form of favorites from many new users.\\

In summary, we see that both preferential creation and preferential reception are occurring in the social interaction layer. There are positive correlations between the number of favorites created/received by users and their probability of creating/receiving new favorites in future. There is also a positive correlation between the number of users favorited by a user and his/her probability of favoriting new users in future. Further, there is a positive correlation between the number of users who favorite a user and the probability of the user receiving favorites from new users in future. However, these positive correlations are not sufficient to claim that a specific mechanism such as the BA model is the cause for the preferential attachment we observe in the social interaction layer.\\

\subsection{Reciprocation}
Reciprocation is a network growth phenomenon in which the creation of a link from one node to another causes the creation of a link in the opposite or reverse direction. The inherent directionality in reciprocation makes it valid only in the case of directed graphs. We wanted to see if favorite markings are reciprocated: if user A favorites a photo uploaded by user B, does that favorite cause user B to later favorite a photo uploaded by user A?\\

\begin{figure}
\centering
\includegraphics[scale=1, width=0.5\textwidth]{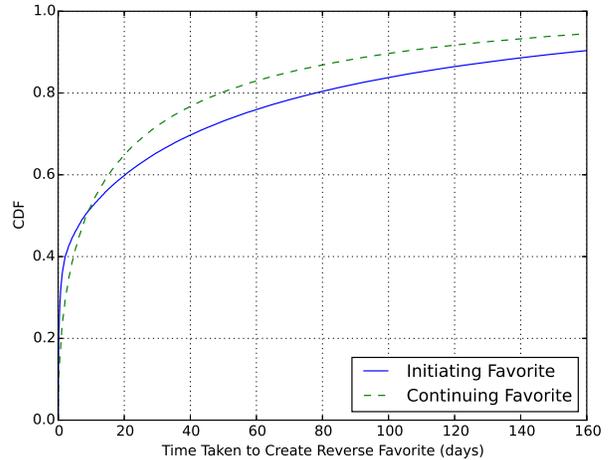}
\caption{CDF of time taken to create a reverse favorite after receiving a favorite. Flickr shows evidence of favorite reciprocation: over 50\% of favorites are reciprocated within 10 days.}
\label{fig:initvscontinuingfav}
\end{figure}

Since it is not possible to know whether a favorite caused the creation of the reverse favorite (the user IDs are anonymized), we look at how long it takes for a reverse favorite to be created after the creation of a favorite. Figure \ref{fig:initvscontinuingfav} shows the cumulative distribution function (CDF) of the time taken for a reverse favorite to be created after a favorite, for both initiating and continuing favorites.

We observe that over 50\% of both kinds of favorites have reverse favorites created within 10 days. Thus, these favorites are more likely to have been the cause for the creation of the reverse favorites. We also observe that for reverse favorites created within 10 days, reverse favorites are created slightly faster in response to initiating favorites than in response to continuing favorites. We hypothesize that this is the case because of a combination of the exploratory nature of Flickr users and homophily: the uploader is more inclined to view the profile of a favoriter who has never favorited his/her photos before (discovery of the favoriter by the uploader) and after the uploader discovers the favoriter, homophily between the uploader and the favoriter leads to a favorite from the uploader to the favoriter (promotion of the favoriter's own photos via the uploader).
While it is less likely that reciprocation occurs after 10 days, we see that for reverse favorites created after 10 days, reverse favorites are created faster in response to continuing favorites than in response to initiating favorites. We hypothesize that this is the case solely because of homophily: a continuing favoriter is more likely to share interests and be a part of the same community as the uploader than a one-off favoriter, leading to the uploader being more likely to favorite his/her favoriter's photos.\\

\section{Multiplex Triangle Closure}
In this section, we examine link formation in the follow layer after a favorite is created in a manner that can be termed as multiplex triangle closure. The role of triangle closure in link formation has long been known to sociologists. Under triangle closure, two people with a mutual friend become friends, thus closing a triangle. Leskovec et al. \cite{leskovec2008microscopic} developed an accurate network evolution model in which each node independently selects destination nodes for creating edges based on the possibility of triangle closure. In a multiplex network however, each of the three edges in a triangle could be in either of the many layers present. For example, in a two-layered multiplex network, 8 different types of triangles are mathematically possible. But not all the types of triangles might make practical sense given the mechanisms of the social networking site which is modeled as a multiplex network. In this paper, we restrict ourselves to one kind of multiplex triangle closure, which we term as \emph{foll-fav-foll}, in which a follow link closes a multiplex triangle after a favorite is created, i.e., we look at those triangles A-B-C in which:\\
\begin{enumerate}[(i)]
\item user A \emph{follows} user B at time $t_{0}$
\item user B \emph{favorites}, at time $t_{1}$, a photo uploaded by user C
\item user A \emph{follows} user C at time $t_{2}$
\item $t_{0} < t_{1} < t_{2}$\\
\end{enumerate}

\begin{figure}
\centering
\includegraphics[scale=1, width=0.5\textwidth]{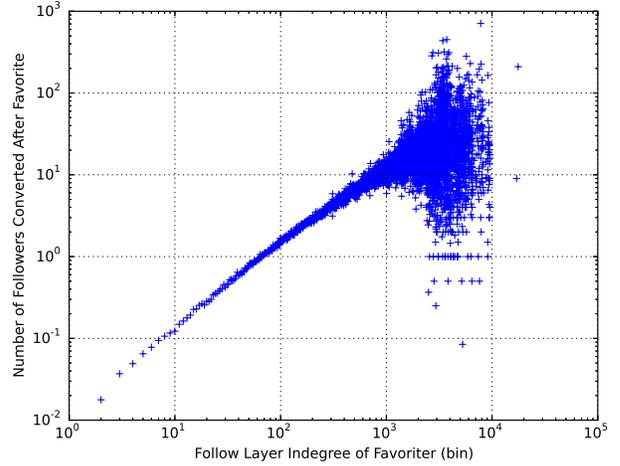}
\caption{Follow indegree of favoriting user (bin) vs. Number of user's followers who become favorite receiver's followers after favorite.}
\label{fig:follsconv}
\end{figure}

Since examining \emph{foll-fav-foll} multiplex triangle closure requires knowing the state of the follow layer before and after a favorite was created, we selected all favorites which were created during the crawl period and examined the creation of \emph{foll-fav-foll} multiplex triangles for each of them. Figure \ref{fig:follsconv} depicts this analysis for all favorites created during the crawl period. We observe that the number of \emph{foll-fav-foll} triangle closures occurring for a favorite after the favorite is created is directly proportional to the number of followers of the favoriting user. In a nutshell, this means that users gain followers from their fan's followers and the gain is proportional to the number of followers of the fan.\\

\section{Related Work}
There has been a long-standing interest in understanding and modeling network growth among researchers and in recent years, social interaction has also become an area of interest, both of which we shall briefly review in this section. Mislove et al. \cite{mislove-2008-flickr} empirically studied the growth of the follow layer in Flickr and found evidence of preferential attachment and reciprocation. They also found a proximity bias in link creation, which they explained by observing that there are few global discovery mechanisms available to Flickr users and that users could primarily explore their neighbourhoods only. Wilson et al. \cite{wilson2009user} proposed the use of interaction graphs as a substitute for social graphs and found that such graphs are better validated by social applications. Viswanath et al. \cite{viswanath2009evolution} studied the evolution of user interaction in the relatively small Facebook New Orleans network and found that the interaction network links became less active as the social links aged. Valafar et al. \cite{valafar2009beyond} studied fan-owner interactions in Flickr and found that a small number of users in the friendship graph are responsible for most interactions. Lipczak et al. \cite{lipczak2013analyzing} analyzed favorites data in Flickr and found that users tend to favorite photos that have been uploaded recently by their friends and that individual favoriting actions tends to occur in bursts. Yang et al. \cite{yang2010understanding} studied the influence of factors such as user, tweet and time on retweeting behavior in Twitter. Lee et al. \cite{lee2010faving} compared favoriting reciprocity in Flickr and Twitter and found significant differences, which they postulated could be due to factors such as the kind of users and the type of content shared in these networks. Macskassy \cite{macskassy2012study} studied social interactions in Twitter on several dimensions, including frequency of interactions and how the interactions are spread across different people. Yang et al. \cite{yang2013unfolding} recently presented a model for the co-evolution of link formation and user interaction in a Chinese social network similar to Facebook. We further contribute to the growing body of knowledge about social interaction by constructing detailed temporal network snapshots using data about favorites and social links and not only examining the growth of the interaction network but also the multiplex effect of interaction on social link formation.\\

\section{Conclusion \& Future Work}
In this paper, we used a first-principles approach to study social interaction in the form of favoriting in Flickr. We found that favoriting of both kinds: photos and users, is well-described by preferential creation and preferential reception. We also found that most favorites are reciprocated within 10 days if at all they are reciprocated and we observed a difference in reciprocation times for initiating and continuing favorites. We examined \emph{foll-fav-foll} multiplex triangle closure, i.e., when user A is a follower of user B and user B favorites a photo uploaded by user C, the creation of a follow link later from user A to user C results in the closure of a multiplex triangle. We found that the number of \emph{foll-fav-foll} multiplex triangle closures occurring for a favorite after it is created is directly proportional to the number of followers of the favoriting user. In summary, our work contributes new insights about favoriting in Flickr and is an important step towards the development of new network growth models that account for the role of social interactions in user discovery and follow link formation.\\

A better understanding of the growth of the social interaction layer can be obtained by exploring the sociological factors behind favoriting a photo, which can be done by conducting user studies and surveys. The occurrence of social interaction after link formation also needs to be considered for the growth of the social interaction layer. The reciprocation of favorites with link formation is also a possibility that needs to be explored given the proximity bias in link creation observed in Flickr. Coupled with knowledge from related work, the empirical observations in this paper and those resulting from pursuing these avenues would serve as a focal point for building an accurate multiplex network growth model for Flickr and other such social networks, which we leave for future work.\\

\section*{Acknowledgment}
The authors would like to thank Alan Mislove for graciously sharing a partly non-anonymized version of the Flickr data set described in \cite{social-cascade-www09}.\\



\bibliographystyle{IEEEtranS}
\bibliography{sigproc}
%



\end{document}